\documentclass[aps,prl,twocolumn,superscriptaddress,floatfix,showpacs]{revtex4}

\usepackage{graphicx}

\usepackage[intlimits]{amsmath}
\usepackage{amscd}
\usepackage{amssymb}
\usepackage{amsfonts}
\usepackage{amsthm}

\usepackage{psfrag}
 
\newcommand{\myfigref}[1]{\mbox{Fig.\,\ref{#1}}}

\begin{document}


\title{Wide-angle perfect absorber/thermal emitter in the THz regime}


\author{Marcus Diem}
\email[]{diem@ameslab.gov}
\affiliation{Ames Laboratory and Department of Physics and Astronomy, Iowa State University, Ames, IA, 50011}
\author{Thomas Koschny}
\author{C. M. Soukoulis}
\affiliation{Ames Laboratory and Department of Physics and Astronomy, Iowa State University, Ames, IA, 50011}
\affiliation{Institute of Electronic Structure (IESL) and Laser, Foundation for Research Technology Hellas (FORTH) and
Department of Material Science and Technology, University of Crete, 71110 Heraklion, Crete, Greece}


\date{\today}

\begin{abstract}
We show that a perfect absorber/thermal emitter exhibiting
an absorption peak of 99.9\% can be achieved in metallic nanostructures
that can be easily fabricated. The very high absorption is
maintained for large angles with a minimal shift in the center frequency
and can be tuned throughout the visible and near-infrared regime by
scaling the nanostructure dimensions. The stability of the spectral
features at high temperatures is tested by simulations using
a range of material parameters.
\end{abstract}

\pacs{44.40.+a,78.20.Ci}


\maketitle




%

Since the beginning of the last century it is known that a perfect thermal emitter
follows Planck's law of black-body radiation\cite{Planck}.
Realistic structures, however, generally do not follow Planck's law, but exhibit
a smaller emission. The properties of these emitters strongly depend on the materials and their shapes.
From the absorption spectra of a structure the emission properties can be deduced since Kirchhoff's law
directly relates the absorption with the emissivity. The emission is then determined by multiplying the
emissivity with the black-body radiation spectrum. Using Photonic Crystals\cite{flem02,chan:036615,han:053906}
it has been shown that this approach is also valid for periodically structured materials.
For a number of applications such as thermo-photovoltaic converters, it is necessary
to control the spectral properties to achieve, \mbox{e.\,g.}, selective emitters in a narrow
frequency band corresponding to the band gap of solar cells\cite{sai:3399,Sai05}. 
In the case of structured metallic surfaces, the changes in the emission spectra are based
on surface waves coupled to the external radiation through the periodic
surface.\cite{Gref02,laroche:123903,Laroche:05}
Alternatively, microcavity resonances can also be used to create narrow band thermal
radiation.\cite{maruyama:1393,celanovic:075127}
Unfortunately, most of the recent designs\cite{laroche:123903,Tao:08}
for perfect absorbers/emitters only
work for one incident angle and one polarization. So, there is a need for wide-angle perfect absorber/emitter
nanostructures.
In this paper, we suggest a structure which exhibits a large absorption in the THz regime for 
a wide range of angles with respect to the surface. We show that the absorption characteristics
are maintained even if the uncertainties in the estimated changes in the material parameters, due to high temperatures,
are considered. The proposed structure can be easily manufactured with today's planar micro-fabrication
techniques. We also comment on the impact of deviations in the geometrical parameters caused by fabricational tolerances.
The small size of the structure, in comparison to the wavelength together with the relatively straightforward fabrication,
allows for easy integration into various devices, such as perfect thermal emitters, perfect absorbers, bolometers, and very
effective light extraction LEDs.

\begin{figure} [ht]
\centering\mbox{}\hfill
\psfrag{myLattice}[l][l]{\Large $\mathrm{a}$}
\psfrag{myTheta}[l][l]{\Large $\theta$}
\psfrag{myWireWidth}[l][l]{\Large$\mathrm{W}_w$}
\psfrag{myWireHeight}[l][l]{\Large$\mathrm{W}_t$}
\psfrag{myDielectricHeight}[l][l]{\Large$\mathrm{D}_t$}
\includegraphics[width=7.5cm]{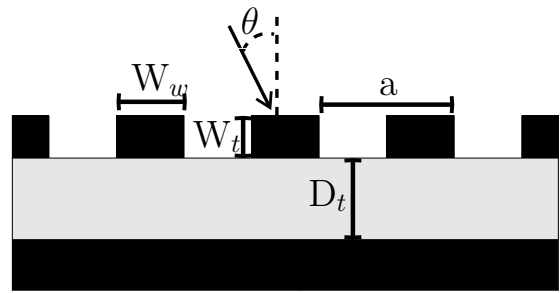}\hfill\mbox{}
\caption{Definition of the structure and parameters. An incoming plane wave with an angle $\theta$ to
the surface normal is considered. The structure consists of an array (lattice constant
$\mathrm{a}\,$=$\,2\mu m$) of Tungsten wires 
($\mathrm{W}_t\,$=$\,0.2\mu m$, $\mathrm{W}_w\,$=$\,0.4\mu m$) on top of a Silicon-Nitride substrate ($\mathrm{D}_t\,$=$\,0.65\mu m$).
The Tungsten plate at the bottom must be thicker than $200nm$. For material parameters see text.}
\label{fig:structure}
\end{figure}

The suggested structure is shown in \myfigref{fig:structure}. It consists of a metal back-plate (black) with a 
thickness larger than $200nm$. This is much larger than the typical skin depth in the THz regime
and avoids transmission through the structure. In this case the reflection is the only factor limiting
the absorption. The thickness of the back-plate can be adjusted to the specific needs of the
final application, \mbox{e.\,g.},  to obtain good heat transport to sensors or to obtain a better stability.
On top of the metal plate a spacer layer of Silicon-Nitride SiN is deposited with a thickness $\mathrm{D}_t$.
The structure is terminated by an array of metallic stripes with a rectangular cross-section. Their
arrangement is described by a lattice constant, $\mathrm{a}$, and their shape is given by a width,
$\mathrm{W}_w$, and a thickness, $\mathrm{W}_t$. In this setup a strong resonance with a large field
enhancement in the dielectric spacer layer and in between the stripes can be obtained, as will be shown later.
Adjusting the size of the metal stripes on the top, the coupling to this resonance can be tuned
and the reflection can be minimized.

Due to the scalability of Maxwell's Equations, in principle, the structure can be
simulated using dimensionless units by dividing all sizes by the
lattice constant and using $\omega'=\mathrm{a}/\lambda$ as
frequency. However, the Drude model used to describe the metal
requires frequencies in $\mathrm{THz}$ and therefore the lattice constant
must be assigned in the simulation. If a shift in the frequencies of the spectral features
by adjusting the lattice constant is intended, a new simulation must be done since changes
in the dielectric constant would not be considered.

In the simulation frequency-dependent material parameters are required.
We calculate them using standard methods and adjust their values to take into
consideration the high temperatures. The Tungsten parts (plate and stripes) are
described by a Drude model
\begin{equation*}
\epsilon = 1 - \frac{\omega_p^2}{\omega\left(\omega + \mathrm{i} \omega_c\right)}
\end{equation*}
with the plasma frequency $\omega_p=1448\mathrm{THz}$ and the collision frequency $\omega_c=13\mathrm{THz}$
at room temperature.\cite{Ordal} In order to account for the higher
temperatures, we use an increased value for the collision frequency. Since the resistivity of Tungsten increases linearly
with temperature, a linear dependence of the collision frequency is assumed, leading to an estimated increase
of the collision frequency by a factor of 3--5 ($\omega_c=50\mathrm{THz}$). The plasma frequency is assumed
to be constant. The simulations are repeated for several values of the collision
frequency ($13,50,100\mathrm{THz}$) to ensure the stability of the emission spectra in a wide range.

The index of SiN depends on the actual fabrication process\cite{piccirillo:3910,matres14,ao38}
and shows a strong wavelength dependence below $1\mu m$, but is roughly constant for longer wavelengths until
about $8\mu m$.\cite{IEEE11} For our simulation we use experimental values for room temperature
obtained from Sandia National Lab\cite{KadySiN} and fitted a Cauchy model for dielectrics in the infrared to calculate
the required values. In order to account for temperature increase, we assume a linear temperature dependence
as found, \mbox{e.\,g.}, in Silica\cite{apllOpt26} and add a constant value ($+0.1$) to the imaginary
part of the experimental data. Again, we perform several simulations with different values of the imaginary parts
of SiN (0.05, 0.1, 0.15, 0.2) for comparison.

For the simulations we use our own implementation of a Fourier-Modal method with a scattering matrix approach,
also known as Rigorous-Coupled Wave Analysis (RCWA).\cite{MG81,WC99} Special care is taken in use of the
correct Fourier-factorization rules to ensure a fast convergence.\cite{LiMath96}
This approach
assumes incoming plane waves onto a periodic surface defined by the dimensionless frequency
$\omega' = \omega a/2\pi c=\mathrm{a}/\lambda$ and the angle $\theta$ to the
surface normal (\myfigref{fig:structure}). For the material parameters the frequency is calculated
using a fixed lattice constant of $\mathrm{a}\,$=$\,2\mu m$. The absorption is determined by subtracting
the total reflection of all diffraction orders given by the individual Poynting vectors. The transmission
is on the order of the numerical error. By Kirchoff's Law the absorption equals emissivity
and the emission can be calculated by multiplying the absorption with the black-body radiation, even in the case of
structured materials.\cite{chan:036615}

To confirm the results and to study the field distributions/energy flow and
resistive heating, we use the commercially-available finite-element-based program, Multiphysics, by Comsol with
periodic boundary conditions in the direction parallel to the surface. In this case the absorption is
determined by $A=1-|S_{21}|^2-|S_{11}|^2$.
This program is also used to determine possible eigenmodes and their corresponding complex 
eigenfrequencies. In this simulation, the material parameters are set to the values at the resonance frequency.

\begin{figure}[ht]
\centering\mbox{}\hfill
\includegraphics[width=7.5cm]{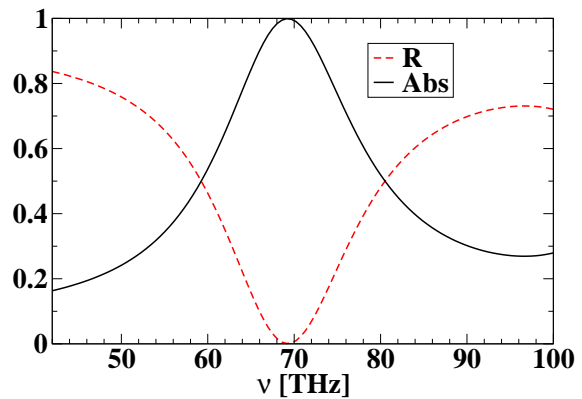}\hfill\mbox{}
\caption{\label{fig:proto}(Color online) Absorption and reflection for perpendicular incidence. The Tungsten back-plate is thicker than
the skin depth and does not allow for transmission. The absorption reaches more than $99.9\%$
for \mbox{$\nu=69.24\mathrm{THz}$} \mbox{($\lambda=4,3\mu m$)}.}
\end{figure}

For wide-angle absorption in a small frequency window, the center frequency of the absorption peak should not shift for
oblique incoming plane waves at different angles. It was shown before that Tungsten microcavities can exhibit
a weak angular dependence,\cite{sai:1685,Kus04} however, the angular range and maximum absorption both are smaller
than in our case.
The additional dielectric layer offers the possibility to optimize the cavity geometry and the coupling
to it lead to a substantial improvement. We find that the most important criteria for obtaining a very weak angular
dependence is to avoid coupling to propagating $-1^{\mathrm{st}}$ diffraction orders. In the simulations we choose
a lattice constant of $\mathrm{a}\,$=$\,2\mu m$, so that the first negative diffraction order at large angles ($\theta\approx90^\circ$)
is given by $\lambda/\mathrm{a}\approx 2$ corresponding to $75\mathrm{THz}$. Any absorption peak at higher
frequencies couples to several reflected diffraction orders and shows a strong angular dispersion at the corresponding
frequency/angle combination. For a structure with the parameters given in \myfigref{fig:structure}, we find
a strong absorption peak for perpendicular incidence fulfilling the above criteria. The corresponding reflection
and absorption spectra are shown in \myfigref{fig:proto}.

\begin{figure}[ht]
\centering\mbox{}\hfill
\includegraphics[width=7.5cm]{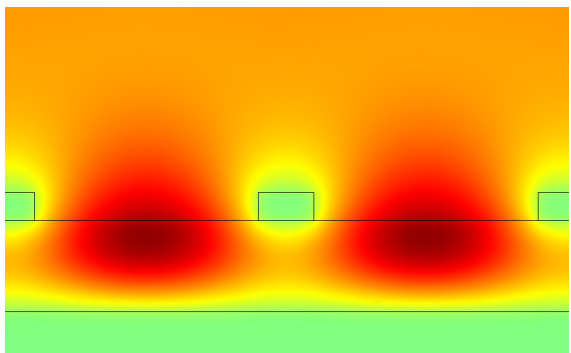}\hfill\mbox{}
\caption{(Color online) Magnitude of the z-component of the electric field at the resonance. The linear scale
ranges ranges from 0.0 (light green) to 25000 (dark red). The maximum of the field is located between the metal stripes
in the dielectric.}
\label{fig:field}
\end{figure}

The obtained absorption peak has a center frequency of $\nu_0\,$=$\,69.24\mathrm{THz}$ with a half-width at half-maximum
of $10.27\mathrm{THz}$ corresponding to 14.8\% of the center frequency. Simulations with Multiphysics
showed, the field is able to couple into the dielectric layer in contrast to regions outside the resonance.
A strong enhancement of the electric field can be found in the region between the metallic stripes mostly located
in the dielectric (\myfigref{fig:field}). Numerically, we find an eigenmode showing the same field pattern at an
eigenfrequency of $70.05\mathrm{THz}$ (real part of the complex eigenfrequency). In this simulation the material parameters are fixed to the
values at $\nu_0$. Changing the thickness of the metal plate or the air space in front of the structure changes the eigenfrequency
by less than $0.01\mathrm{THz}$, once the air part in front of the structure is chosen large enough. Although losses in the dielectric layer
occur due to the imaginary part, the energy flow described by the Poynting vector reveals that the main absorption
takes place at the surface of metallic back-plate and the rear side of the metal stripes.

\begin{figure}[ht]
\centering\mbox{}\hfill
\includegraphics[width=7.9cm]{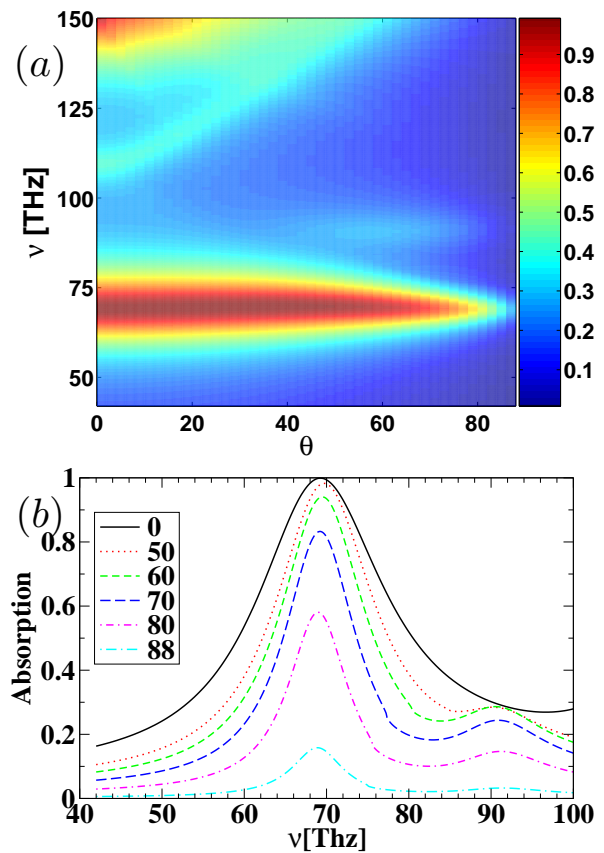}\hfill\mbox{}
\caption{(Color online) (a) Angular peak dispersion of the absorption peak. The additional peaks around 110 and 150THz
show a strong dispersion although they are caused by cavity modes as well. (b) For $0$ to $40^{\circ}$
the absorption is above $99.7\%$. Even at $70^{\circ}$ ($80^\circ$) still 83.3\% (58.1\%) of
the incoming energy is absorbed.}
\label{fig:angle}
\end{figure}

If the structure is used as a perfect absorber, \mbox{e.\,g.}, in sensor applications, it is important to absorb
as much of the radiation as possible, independent of the
direction of incidence. In \myfigref{fig:angle} we show the angular dispersion of the peak together with
the spectra for several angles. For angles up to $40^\circ$ the peak becomes slightly narrower and
the center frequency increases by $0.5\mathrm{THz}$, but the maximum absorption is above 99.7\%. For larger
angles the peak starts to drop, but is still above 80\%, even for waves impinging on the structure with
an angle of $\theta=70^\circ$. In \myfigref{fig:angle}(a),
two additional absorption peaks at higher frequencies are visible. Both peaks are caused by resonant
modes in the dielectric as well, but their angular dependence shows a very strong dispersion.
If the parameters are chosen so that the lowest peak has a higher center frequency (above 75THz for the given
lattice constant), it shows a weak angular dispersion until the $-1^{\mathrm{st}}$ starts propagating and
then the absorption vanishes quickly.

\begin{figure}[ht]
\centering\mbox{}\hfill
\includegraphics[width=7.5cm]{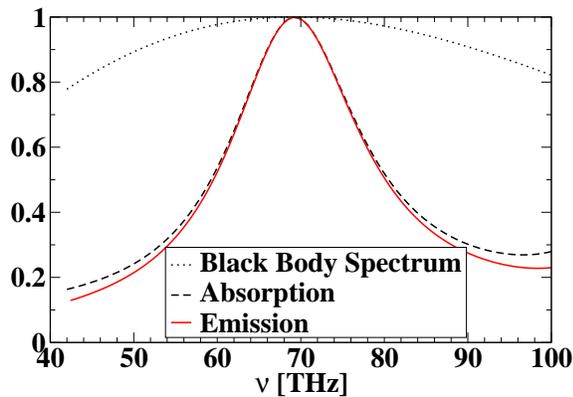}\hfill\mbox{}
\caption{(Color online) Black-body radiation (dotted), absorption (dashed), and emission (solid) of the structure,
if the temperature is adjusted so the maxima of the black-body spectra and the absorption peak coincide using Wien's law (T=1176K).
Emission and absorption are roughly the same, due to high value of the black-body radiation.}
\label{fig:emission}
\end{figure}

Finally, we compare the emission properties of our structure with a regular black-body emitter
in \myfigref{fig:emission}. We plot the black-body emission using Planck's law at
a temperature of 1176K, so the maximum of the emission corresponds to the center frequency
of the absorption peak. The emission of the structure given by the product of the
emissivity and the black-body radiation is also plotted in \myfigref{fig:emission} as a solid line. 
The combined effect of the two aspects suppresses the sides of the peak even stronger,
so a selective emission is achieved.

For these high temperatures, the material parameters are not known exactly and must be estimated.
Since estimates are always a possible source of errors, we ensured that the spectral properties exist
in a wide range of parameters. We tested the results for different adjustments in the collision frequency
of Tungsten ($\omega_c\,$=13, 50, 100THz) and the imaginary part of the SiN dielectric (0.05, 0.1, 0.15, 0.2).
For all combinations of these parameters, the absorption reaches 
values of 93\% or more, except for the case with $\omega_c=13\mathrm{THz}$ and $\mathrm{Im} \epsilon\,$=$\,0.05$
for which only 82\% are achieved. The center frequency of the absorption peak remains unchanged. However,
an increase in the imaginary part of SiN leads to an broadening of it. In general, an increase
in the collision frequency allows for a wider range of possible adjustments in the imaginary
part of the SiN.

Although we do not present the data in this paper, we also conducted several simulations to determine the
sensitivity of the absorption peak to fabricational tolerances. In this study, we varied all 3 parameters
($\mathrm{W}_w$, $\mathrm{W}_t$, $\mathrm{D}_t$) by $\pm50nm$ individually and combined. In all cases an
absorption of more than 98,8\% was obtained. Wider (narrower) stripes lead to a small increase (decrease) in
the frequency of less than $1\mathrm{THz}$. For thicker stripes, the coupling became weaker and the
absorption dropped by 1\%. Changes in the thickness of the dielectric layer caused the strongest change
in the absorption spectra, shifting the center frequency of the peak by about $\pm4\mathrm{THz}$. Since
the field was concentrated mainly in this area, deviations in the spacer thickness changed the effective
size of the cavity and, therefore, the frequency of the cavity mode. This makes it important to control the
thickness of the dielectric layer very precisely in the fabrication process.

In summary, we presented a design for a perfect absorber/thermal emitter with a very high
absorption peak over a wide range of angles. We demonstrated that the absorption peak is  
stable with respect to the estimated changes in the material parameters at
high temperatures. We also studied the impact of deviations in the geometry caused
by fabricational tolerances. From this it can be expected that the spectral
features are also present in realistic samples. Since the structure can be built with planar
fabrication techniques, it offers an interesting approach to wide angle perfect absorbers/emitters.
First studies using a lattice constant of several $mm$ showed that it is possible to obtain
similar effects also for frequencies in the GHz range. Due to the wide angular absorption this
can offer a new way to avoid reflections in microwave experiments with an easy-to-build structure
on length-scales smaller than the wavelength.

M.D. gratefully acknowledges financial support from the Alexander-von-Humboldt 
Foundation (Feodor-Lynen Program).
Work at Ames Laboratory was supported by the Department of Energy (Basic Energy Sciences)
under contract No. DE-AC02-07CH11358. This work was partially supported by the office
of Naval Research (Award No. N0014-07-1-0359).


\end{document}